# Magnetic interactions in the $S$ = 1/2 square-lattice antiferromagnets Ba$_2$CuTeO$_6$ and Ba$_2$CuWO$_6$: parent phases of a possible spin liquid

Otto Mustonen,[*ab] Sami Vasala,[cd] Heather Mutch,[b] Chris I. Thomas,[a] Gavin B. G. Stenning,[e] Elisa Baggio-Saitovitch,[c] Edmund J. Cussen[b] and Maarit Karppinen[*a]

a. Department of Chemistry and Materials Science, Aalto University, FI-00076 Espoo, Finland. E-mail: maarit.karppinen@aalto.fi and ohj.mustonen@gmail.com
b. Department of Materials Science and Engineering, University of Sheffield, Mappin Street, Sheffield S1 3JD, United Kingdom.
c. Centro Brasileiro de Pesquisas Físicas (CBPF), Rua Dr Xavier Sigaud 150, Urca, Rio de Janeiro, 22290-180, Brazil.
d. Technische Universität Darmstadt, Institut für Materialwissenschaft, Fachgebiet Materialdesign durch Synthese, Alarich-Weiss-Straße 2, 64287 Darmstadt, Germany
e. ISIS Neutron and Muon Source, Rutherford Appleton Laboratory, Harwell Science and Innovation Campus, Didcot, OX11 0QX, United Kingdom.

**The isostructural double perovskites Ba$_2$CuTeO$_6$ and Ba$_2$CuWO$_6$ are shown by theory and experiment to be frustrated square-lattice antiferromagnets with opposing dominant magnetic interactions. This is driven by differences in orbital hybridisation of Te$^{6+}$ and W$^{6+}$. A spin-liquid-like ground state is predicted for Ba$_2$Cu(Te$_{1-x}$W$_x$)O$_6$ solid solution similar to recent observations in Sr$_2$Cu(Te$_{1-x}$W$_x$)O$_6$.**

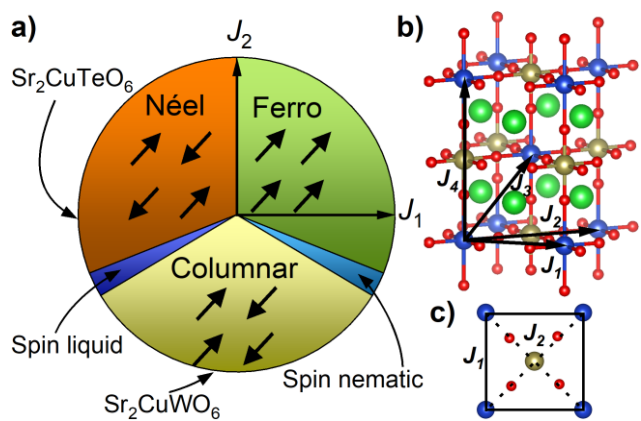

Magnetic frustration can stabilise novel quantum ground states such as quantum spin liquids or valence bond solids.[1] Frustration occurs when not all of the magnetic interactions in a material can be satisfied simultaneously as a result of lattice geometry or competing interactions. We have recently shown that a quantum-spin-liquid-like state forms in the double perovskite solid solution Sr$_2$Cu(Te$_{1-x}$W$_x$)O$_6$ with a square lattice of Cu$^{2+}$ ($3d^9$, $S$ = 1/2) cations.[2,3] This was the first observation of a spin-liquid-like state in a square-lattice compound after 30 years of theoretical predictions.[4–7]

The parent compounds Sr$_2$CuTeO$_6$ and Sr$_2$CuWO$_6$ are frustrated square-lattice (FSL) antiferromagnets.[8–12] The FSL model (Fig. 1) has two interactions: nearest-neighbour $J_1$ interaction (side) and next-nearest-neighbour $J_2$ interaction (diagonal). Dominant antiferromagnetic $J_1$ leads to Néel type antiferromagnetic order and dominant $J_2$ leads to columnar magnetic order. Magnetic frustration arises from the competition of $J_1$ and $J_2$, and a quantum spin liquid state has been predicted for $J_2/J_1$ = 0.5 where frustration is maximised.[4–7]

Sr$_2$CuTeO$_6$ and Sr$_2$CuWO$_6$ are the first known isostructural FSL systems with different dominant interactions and magnetic structures: dominant $J_1$ and Néel order for Sr$_2$CuTeO$_6$ and dominant $J_2$ and columnar order for Sr$_2$CuWO$_6$ respectively.[8,9] The two compounds have a tetragonal $I4/m$ double perovskite structure with nearly identical bond distances and angles.[10,12] The magnetism becomes highly two-dimensional as a result of a Jahn-Teller distortion as the only unoccupied Cu orbital $3d_{x^2-y^2}$ is in the ab square plane. The major differences in dominant magnetic interactions are due to the diamagnetic Te$^{6+}$ $d^{10}$ and W$^{6+}$ $d^0$ cations located in the middle of the Cu$^{2+}$ square (Fig. 1c), which hybridise differently with O 2p allowing different superexchange paths between the Cu$^{2+}$ cations.[13,14] The spin-liquid-like ground state forms when these two perovskites are mixed into a Sr$_2$Cu(Te$_{1-x}$W$_x$)O$_6$ solid solution.[2,3,15] Muon spin relaxation experiments revealed the absence of magnetic order or static magnetism in a wide composition range of $x$ = 0.1-0.6.[2,3] The specific heat displays $T$-linear behaviour suggesting gapless excitations in a similar composition range.[2,3,15] The ground state has been proposed to be a random-singlet state with a disordered arrangement of non-magnetic valence bond singlets.[16]

Motivated by these exciting findings in the Sr$_2$Cu(Te$_{1-x}$W$_x$)O$_6$ system, we have investigated the magnetic interactions of the isostructural barium analogues Ba$_2$CuTeO$_6$ and Ba$_2$CuWO$_6$. Ba$_2$CuWO$_6$ is known to have columnar magnetic order,[17,18] but little is known about Ba$_2$CuTeO$_6$ as the perovskite phase requires high pressures to synthesise.[19] Here we use density functional theory (DFT) calculations and high-temperature series expansion (HTSE) fitting of experimental susceptibility data to show that these compounds are FSL antiferromagnets with opposite dominant interactions similar to Sr$_2$CuTeO$_6$ and Sr$_2$CuWO$_6$. We predict a quantum-spin-liquid-like

**Fig. 1. a)** Phase diagram of the frustrated square-lattice model. Antiferromagnetic (negative) $J_1$ stabilises Néel order and $J_2$ columnar order respectively. A spin liquid state has been predicted for the Néel–columnar boundary at $J_2/J_1$ = 0.5 where magnetic frustration is maximised. **b)** The double perovskite structure of (Ba,Sr)$_2$Cu(Te,W)O$_6$. $J_1$ and $J_2$ are the in-plane interactions of the FSL model, whereas $J_3$ and $J_4$ are out-of-plane interactions. The blue, dark yellow, red and green spheres represent Cu, Te/W, O and Ba/Sr, respectively. **c)** The Cu$^{2+}$ square in the $ab$ plane with $J_1$ and $J_2$ interactions.

state in $Ba_2Cu(Te_{1-x}W_x)O_6$ with strong antiferromagnetic interactions.

Magnetic interactions and electronic structure in $Ba_2CuTeO_6$ and $Ba_2CuWO_6$ were calculated using the DFT+$U$ framework, where an on-site Coulomb repulsion term $U$ was used to model electron correlation effects of localised Cu 3$d$ orbitals. Interactions up to the fourth-nearest neighbour were evaluated, see Fig. 1b. $J_1$ and $J_2$ are the square plane interactions of the FSL model, and $J_3$ and $J_4$ are additional out-of-plane interactions. Energies of different spin configurations were mapped onto a Heisenberg Hamiltonian to obtain $J_1$-$J_4$. We have previously shown this approach works well for $Sr_2CuWO_6$.[9] The $J_1$ and $J_2$ interactions were also determined from experimental magnetic susceptibility data using high-temperature series expansion fitting. $Ba_2CuTeO_6$ was prepared by high-pressure synthesis and $Ba_2CuWO_6$ by conventional solid state synthesis. Details of the DFT calculations, sample synthesis and characterisation are available in the electronic supporting information (ESI).

**Table 1.** Exchange constants of Ba2CuTeO6 and Ba2CuWO6 obtained by density functional theory using different on-site Coulomb U terms and by high-temperature series expansion fitting of magnetic susceptibility data. Negative (positive) values correspond to antiferromagnetic (ferromagnetic) interactions.

| $Ba_2CuTeO_6$ | $U$ = 7 eV | $U$ = 8 eV | $U$ = 9 eV | HTSE |
|---|---|---|---|---|
| $J_1$ (meV) | -23.65 | -20.22 | -17.22 | -16.54(3) |
| $J_2$ (meV) | 0.13 | 0.23 | 0.06 | -0.04(3) |
| $J_3$ (meV) | 1.28 | 0.83 | 0.67 | - |
| $J_4$ (meV) | -0.30 | 0.01 | 0.05 | - |
| $J_2/J_1$ | -0.01 | -0.01 | -0.003 | 0.002 |
| $Ba_2CuWO_6$ | $U$ = 7 eV | $U$ = 8 eV | $U$ = 9 eV | HTSE |
| $J_1$ (meV) | -1.25 | -1.17 | -1.27 | 0.2(9) |
| $J_2$ (meV) | -14.71 | -11.94 | -9.56 | -10.0(1) |
| $J_3$ (meV) | 0.05 | -0.01 | 0.01 | - |
| $J_4$ (meV) | 0.03 | 0.37 | 0.02 | - |
| $J_2/J_1$ | 11.79 | 10.18 | 7.55 | -50* |

*significant uncertainty in this value due to error in J1

The calculated magnetic interactions of $Ba_2CuTeO_6$ and $Ba_2CuWO_6$ are presented in Table 1. The calculated values depend on the Coulomb $U$ term as is typical with DFT+$U$, but the same trends are observed for reasonable values of $U$. Despite being isostructural, the magnetic interactions in $Ba_2CuTeO_6$ and $Ba_2CuWO_6$ are very different. $Ba_2CuTeO_6$ has a very dominant antiferromagnetic $J_1$ interaction with weak $J_2$, $J_3$ and $J_4$ interactions. It is a near-ideal FSL Néel antiferromagnet. $Ba_2CuWO_6$, in contrast, has a dominant antiferromagnetic $J_2$ interaction slightly frustrated by an antiferromagnetic $J_1$ interaction with negligible $J_3$ and $J_4$ interactions. The strong $J_2$ interaction is consistent with the known columnar magnetic structure of this compound.[18] Due to the weakness of the out-of-plane $J_3$ and $J_4$ interactions, magnetism in both compounds is highly two-dimensional and well described by the FSL model.

The significant differences in the magnetic interactions of $Ba_2CuTeO_6$ and $Ba_2CuWO_6$ can be explained by their electronic structures. We have plotted total and partial densities of states for both compounds in Fig. 2. $Ba_2CuTeO_6$ and $Ba_2CuWO_6$ are antiferromagnetic insulators: the band gaps open between the occupied Cu 3$d$ states hybridised with O 2$p$ (valence band) and the unoccupied Cu $3d_{x^2-y^2}$ states hybridised with O 2$p$ (conduction band). In $Ba_2CuWO_6$ the conduction band is further hybridised with unoccupied W 5$d$ states. The W 5$d$ states also hybridise with the Cu 3$d$/O 2$p$ states in the valence band, which allows a 180° Cu-O-W-O-Cu superexchange pathway resulting in a strong antiferromagnetic $J_2$ interaction. This hybridisation does not occur in $Ba_2CuTeO_6$ and therefore $J_2$ is negligible. In $Ba_2CuTeO_6$ the Te 5$p$ states hybridise to a lesser degree with the Cu 3$d$/O 2$p$ states in the conduction band, which could explain the strong antiferromagnetic $J_1$ interaction. However, the role of Te in the $J_1$ superexchange in $Sr_2CuTeO_6$ is under debate.[8,13] Overall, the electronic structures of $Ba_2CuTeO_6$ and $Ba_2CuWO_6$ are similar to their strontium analogues $Sr_2CuTeO_6$ and $Sr_2CuWO_6$, and the differences in magnetic interactions are driven by the same orbital hybridisation mechanism.

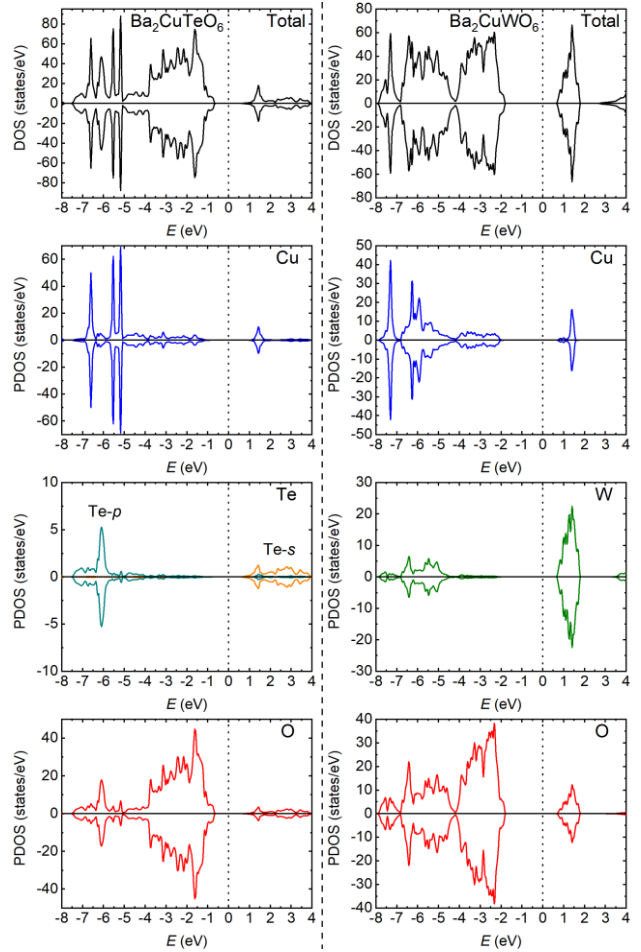

**Fig. 2**. Total and partial density of states plots for $Ba_2CuTeO_6$ (left) and $Ba_2CuWO_6$ (right). Both compounds are antiferromagnetic insulators. The moderate Te 5$p$/5$s$ – O 2$p$ hybridisation and stronger W 5$d$ – O2$p$ hybridisation are seen in the Te/W and O PDOS plots.

The experimental magnetic susceptibilities of synthesised Ba2CuTeO6 and Ba2CuWO6 samples are shown in Fig. 3. The broad maximum observed in the susceptibility is due to the two-dimensional nature of the magnetism in these materials. Our maximum temperature of 400 K was not enough for reliable

Curie-Weiss fits. Previous measurements[19] up to 800 K yielded the Curie-Weiss constants $\Theta_{CW}$ = -400 K for $Ba_2CuTeO_6$ and $\Theta_{CW}$ = -249 K for $Ba_2CuWO_6$ revealing strong antiferromagnetic interactions.

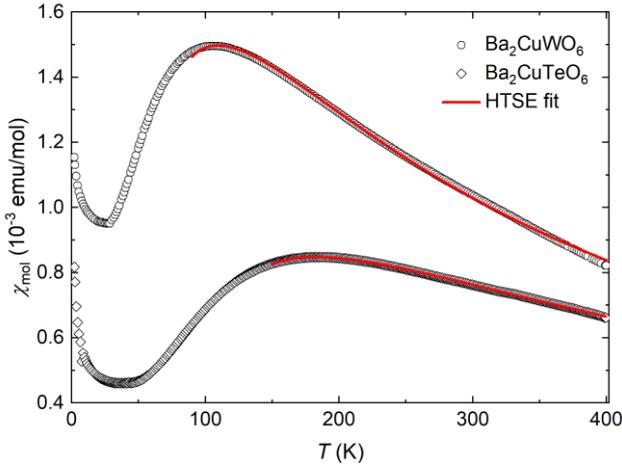

**Fig. 3.** Magnetic susceptibility and high-temperature series expansion fits for $Ba_2CuTeO_6$ and $Ba_2CuWO_6$. Open symbols represent experimental data and the lines are HTSE fits with the parameters $J_1$ = -16.54(3) meV, $J_2$ = -0.04(3) meV, $g$ = 2.20(1) and $J_1$ = 0.2(9) meV, $J_2$ = -10.0(1) meV, $g$ = 2.26(5) for $Ba_2CuTeO_6$ and $Ba_2CuWO_6$, respectively. The ZFC and FC curves overlap and therefore only ZFC data is shown.

The magnetic susceptibilities were fitted to a high-temperature series expansion of the FSL model.[20] The molar magnetic susceptibility $\chi_{mol}$ is given by:

$$\chi_{mol} = \frac{N_A g^2 \mu_B^2}{k_B T} \sum_n \beta^n \sum_m c_{m,n} x^m + \chi_0 \qquad (1)$$

where $g$ is the effective $g$-factor, $\beta$ = -$J_1/k_B$, $x$ = $J_2/J_1$, $\chi_0$ is a temperature independent diamagnetic correction and the coefficients $c_{m,n}$ are from Table I in ref. 20. The model has four parameters: $J_1$, $J_2$, $g$ and $\chi_0$, which were fitted to the experimental data using a least squares method. The model always produces two solutions due to internal symmetry: one with dominant $J_1$ and one with dominant $J_2$.[21] Our DFT calculations allow us to select the correct dominant $J_1$ solution for $Ba_2CuTeO_6$ and the dominant $J_2$ solution for $Ba_2CuWO_6$.

The best fits were obtained with the parameters $J_1$ = 16.54(3) meV, $J_2$ = -0.04(3) meV, $g$ = 2.20(1) for $Ba_2CuTeO_6$ and $J_1$ = 0.2(9) meV, $J_2$ = -10.0(1) meV, $g$ = 2.26(5) for $Ba_2CuWO_6$ in the temperature ranges 150-400 K and 90-400 K, respectively. The fitted exchange constants depend slightly on the minimum temperature used. For both compounds the calculated dominant interaction remains stable in a wide fitting range, but the weaker interaction cannot be accurately quantified. In $Ba_2CuTeO_6$ the sign of $J_2$ changes depending on the fitting range, whereas in $Ba_2CuWO_6$ the error of $J_1$ is much larger than its value. We can conclude, however, that the dominant interaction is much stronger than the weak one in both $Ba_2CuTeO_6$ ($|J_2|/|J_1|$ < 0.02) and $Ba_2CuWO_6$ ($|J_1|/|J_2|$ < 0.12) and that the DFT and HTSE results are in good agreement.

The magnetic properties of $Ba_2CuTeO_6$, $Ba_2CuWO_6$, $Sr_2CuTeO_6$ and $Sr_2CuWO_6$ are summarised in Table 2. Magnetic interactions in $Ba_2CuTeO_6$ and $Ba_2CuWO_6$ are notably stronger than their strontium analogues. This is due to the smaller tilting of the $CuO_6$ octahedra in the barium phases, which leads to stronger orbital overlap as the Cu-O-Te/W angle is closer to 180 degrees.[19] As long-range magnetic order is driven by the weak out-of-plane interactions which are of the same order in all compounds, $Ba_2CuTeO_6$ and $Ba_2CuWO_6$ are even closer to ideal two-dimensional antiferromagnets than their strontium analogues. The transition temperature of $Ba_2CuTeO_6$ is not known, but we predict it to have the highest frustration index $f$ = $\Theta_{CW}/T_N$ of these compounds and the Néel magnetic structure due to the very strong $J_1$ interaction. Magnetic excitations in $Sr_2CuTeO_6$ and $Sr_2CuWO_6$ have been observed at temperatures higher than $2T_N$ driven by the two-dimensional magnetic interactions.[8,9] The stronger in-plane $J_1$ and $J_2$ interactions of the barium phases indicate the excitations survive to even higher temperatures.

**Table 2.** Magnetic properties of $Ba_2CuTeO_6$, $Sr_2CuTeO_6$, $Ba_2CuWO_6$ and $Sr_2CuWO_6$. Exchange interactions $J_1$ and $J_2$ have been obtained by density functional theory (DFT; $U$ = 8 eV), high-temperature series expansion fitting (HTSE) or by inelastic neutron scattering (INS). The data for $Ba_2CuTeO_6$ and $Ba_2CuWO_6$ are from this work unless specified otherwise.

| | $Ba_2CuTeO_6$ | $Sr_2CuTeO_6$ | $Ba_2CuWO_6$ | $Sr_2CuWO_6$ |
|---|---|---|---|---|
| $J_1$ (meV) | -20.22 (DFT) -16.54(3) (HTSE) | -7.18 (INS)[8] | -1.17 (DFT) -0.2(9) (HTSE) | -2.45 (DFT)[9] -1.2 (INS)[9] |
| $J_2$ (meV) | 0.23 (DFT) -0.04(3) (HTSE) | -0.21 (INS)[8] | -11.94 (DFT) -10.0(1) (HTSE) | -8.83 (DFT)[9] -9.5 (INS)[9] |
| $\Theta_{CW}$ (K) | -400[19] | -80[2] | -249[19] | -165[2] |
| $T_N$ (K) | - | 29[10] | 28[18] | 24[12] |
| $f=\Theta_{CW}/T_N$ | - | 2.8 | 8.9 | 6.9 |
| $k$ | [1/2 1/2 $k_z$]* | [1/2 1/2 0][10] | [0 1/2 1/2][18] | [0 1/2 1/2][11] |
| Magnetic order | Néel* | Néel | Columnar | Columnar |

*predicted based on magnetic interactions

Since $Ba_2CuTeO_6$ has a dominant $J_1$ interaction and $Ba_2CuWO_6$ has a dominant $J_2$ interaction, we predict a spin-liquid-like state will form in the $Ba_2Cu(Te_{1-x}W_x)O_6$ solid solution similar to $Sr_2Cu(Te_{1-x}W_x)O_6$. In the $Sr_2Cu(Te_{1-x}W_x)O_6$ system the Néel order is destabilised already at $x$ = 0.1, and spin-liquid-like state exist in the composition region $x$ = 0.1-0.6. Columnar order is observed for $x$ = 0.7-1. Since the $J_1$ interaction of $Ba_2CuTeO_6$ is so strong even compared to $J_2$ in $Ba_2CuWO_6$, we predict the Néel order remains more stable against W substitution. For the same reason, the columnar order near $x$ = 1 is likely to be less stable in $Ba_2Cu(Te_{1-x}W_x)O_6$. The extent of the spin-liquid-like region depends also on disorder, and is difficult to predict just from the properties of the end phases. Finally, the stronger antiferromagnetic interactions in the barium phases indicate that the quantum disordered ground state will remain stable up to higher temperatures.

The previous discussion concerns a double perovskite $Ba_2Cu(Te_{1-x}W_x)O_6$ solid solution, which near $x$ = 0 will require high-pressure synthesis to form. The ambient pressure form of $Ba_2CuTeO_6$ is triclinic with a tolerance factor higher than 1.03.[22] Therefore, a $Ba_2Cu(Te_{1-x}W_x)O_6$ solid solution prepared in ambient pressure will have a triclinic to tetragonal structural change at some composition. Triclinic $Ba_2CuTeO_6$ is a spin ladder system close to a quantum critical point,[23] and we propose Te-for-W substitution could drive the system from magnetic order to a spin singlet state.

In conclusion, we have investigated the magnetic interactions of the tetragonal double perovskites $Ba_2CuTeO_6$ and $Ba_2CuWO_6$ by DFT calculations and by HTSE fitting. Both compounds are well described by the frustrated square-lattice model as out-of-plane interactions are very weak. In $Ba_2CuTeO_6$ the antiferromagnetic nearest-neighbor $J_1$ interaction dominates ($|J_2|/|J_1| < 0.02$), whereas in $Ba_2CuWO_6$ the antiferromagnetic next-nearest neighbor interaction $J_2$ dominates ($|J_1|/|J_2| < 0.12$). The $Ba_2Cu(Te,W)O_6$ system is the second known FSL system where isostructural compounds have opposite magnetic interactions. This is driven by differences in orbital hybridisation of Te $5p/5s$ and W $5d$ with O $2p$. A spin-liquid-like ground state is predicted for the $Ba_2Cu(Te_{1-x}W_x)O_6$ solid solution similar to the recent findings in the $Sr_2Cu(Te_{1-x}W_x)O_6$ system.

The authors wish to acknowledge CSC – IT Center for Science, Finland, for computational resources. The authors are thankful for access to the MPMS3 instrument in the Materials Characterisation Laboratory at the ISIS Muon and Neutron Source. OM, HM and EJC are thankful for funding by the Leverhulme Trust Research Project Grant 2017-109. SV is thankful for the support of the Brazilian funding agencies CNPq (grants no. 150503/2016-4 and 152331/2016-6) and FAPERJ (grant no. 202842/2016). EBS is grateful for financial support by a joint DFG-FAPERJ project DFG Li- 244/12. In addition, EBS acknowledges support from FAPERJ through several grants including Emeritus Professor fellow and CNPq for BPA and corresponding grants.

**Conflicts of interest**

There are no conflicts to declare.

**Notes and references**

# Magnetic interactions in the S = 1/2 square-lattice antiferromagnets Ba$_2$CuTeO$_6$ and Ba$_2$CuWO$_6$

Otto Mustonen,[a,b] Sami Vasala,[c,d] Heather Mutch,[b] Chris I. Thomas,[a] Gavin B. G. Stenning,[e] Elisa Baggio-Saitovitch,[c] Edmund J. Cussen[b] and Maarit Karppinen [*a]

**Electronic Supporting Information**

**Density functional theory calculations**

Density functional theory was used to calculate the magnetic exchange constants in Ba$_2$CuTeO$_6$ and Ba$_2$CuWO$_6$. The calculations were carried out with the full potential linearized augmented plane wave code ELK.[1] We used the generalized gradient approximation functionals by Perdew, Burke and Ernzerhof.[2] Five different spin configurations with 2 × 2 × 1 (1 × 1 × 2) supercells were needed to calculate the exchange constants (Fig. 1.).[3,4] A $k$ point grid of 4 × 4 × 6 (8 × 8 × 3) was used. A plane-wave cutoff of $|G + k|_{max}$ = 8/$R_{MT}$ a.u.$^{-1}$ was used, where $R_{MT}$ was the average muffin tin radius. Electron correlation effects of the localized Cu$^{2+}$ 3$d$ orbitals were included within the DFT+$U$ framework with the on-site coulombic repulsion $U$ and Hund exchange term $I$ as parameters.[5] The on-site coulombic $U$ term was varied from 7 to 9 eV, which are typical values for Cu 3$d$ orbitals. The Hund term $I$ was fixed at 0.9 eV for all calculations.

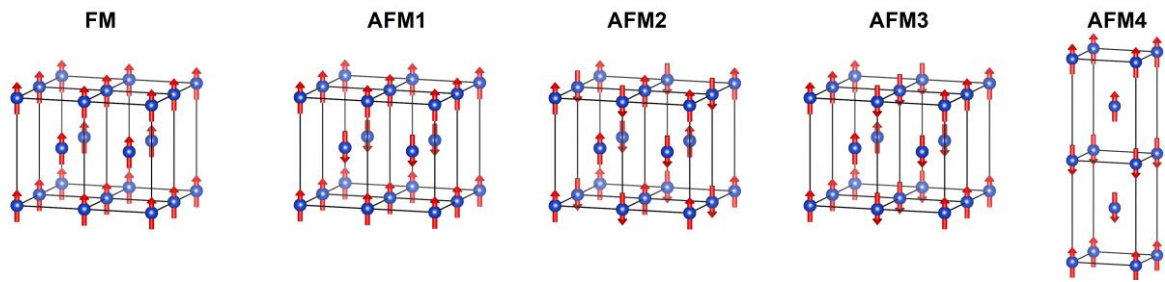

Fig. 1. The five different spin configurations used in the density functional theory calculations. Only the magnetic Cu$^{2+}$ cations and their spins are shown. The energies are calculated in 2 × 2 × 1 (and one 1 × 1 × 2) supercells.

In order to obtain the exchange constants $J_1$-$J_4$ we mapped the energies of the different spin configurations to a simple Heisenberg Hamiltonian:

$$H = -\sum_{i<j} J_{ij} S_i \cdot S_j$$

where $J_{ij}$ is the exchange constant for the interaction between spins $i$ and $j$. The spin configurations are presented in Fig. 1. Using the Hamiltonian, the energies of the spin configurations[3] can be written as:

$$E_{FM} = E_0 + (-4J_1 - 4J_2 - 8J_3 - 2J_4)S^2$$
$$E_{AFM1} = E_0 + (-4J_1 - 4J_2 + 8J_3 - 2J_4)S^2$$

$$E_{AFM2} = E_0 + (4J_1 - 4J_2 - 2J_4)S^2$$
$$E_{AFM3} = E_0 + (4J_2 - 2J_4)S^2$$
$$E_{AFM4} = E_0 + (-4J_1 - 4J_2 + 2J_4)S^2$$

The exchange constants $J_1$-$J_4$ can then be obtained from:[3]

$$J_3 = (E_{AFM1} - E_{FM})/16S^2$$
$$J_1 = (E_{AFM2} - E_{FM} - 8J_3S^2)/8S^2$$
$$J_2 = (E_{AFM3} - E_{FM} - 4J_1S^2 - 8J_3S^2)/8S^2$$
$$J_4 = (E_{AFM4} - E_{FM} - 8J_3S^2)/4S^2$$

The calculated energies and exchange constants for $U$ = 7-9 eV are presented in Table 1.

Table 1. Relative total energies of the different spin configurations of $Ba_2CuTeO_6$ and $Ba_2CuWO_6$ calculated by density functional theory. Energy of the ferromagnetic configuration is set as zero.

|  | $Ba_2CuTeO_6$ | | | $Ba_2CuWO_6$ | | |
| --- | --- | --- | --- | --- | --- | --- |
|  | $U$ = 7 eV | $U$ = 8 eV | $U$ = 9 eV | $U$ = 7 eV | $U$ = 8 eV | $U$ = 9 eV |
| $E_{FM}$ (meV/2f.u.) | 0 | 0 | 0 | 0 | 0 | 0 |
| $E_{AFM1}$ (meV/2f.u.) | 5.12 | 3.33 | 2.67 | 0.22 | -0.04 | 0.04 |
| $E_{AFM2}$ (meV/2f.u.) | -44.74 | -38.78 | -33.11 | -2.39 | -2.37 | -2.51 |
| $E_{AFM3}$ (meV/2f.u.) | -20.82 | -18.10 | -15.77 | -30.56 | -25.08 | -20.36 |
| $E_{AFM4}$ (meV/2f.u.) | 2.26 | 1.67 | 1.39 | 0.14 | 0.35 | 0.04 |
| $J_1$ (meV) | -23.65 | -20.22 | -17.22 | -1.25 | -1.17 | -1.27 |
| $J_2$ (meV) | 0.13 | 0.23 | 0.06 | -14.71 | -11.94 | -9.56 |
| $J_3$ (meV) | 1.28 | 0.83 | 0.67 | 0.05 | -0.01 | 0.01 |
| $J_4$ (meV) | -0.30 | 0.01 | 0.05 | 0.03 | 0.37 | 0.02 |
| $J_2/J_1$ | -0.01 | -0.01 | 0.00 | 11.79 | 10.18 | 7.55 |

**Sample synthesis**

$Ba_2CuWO_6$ and triclinic $Ba_2CuTeO_6$ were prepared using a conventional solid state reaction method from stoichiometric amounts of $BaCO_3$, $CuO$, $WO_3$ and $TeO_2$ (Alpha Aesar ≥99.995). The samples were calcined at 900 °C in air for 12 hours, reground, pelletized and fired twice at 1000 °C in air for 24 hours. Tetragonal double perovskite $Ba_2CuTeO_6$ was prepared from triclinic $Ba_2CuTeO_6$ under high-pressure high-temperature conditions. Sample powder enclosed in a gold capsule was pressed in a cubic-anvil Riken-Seiki high-pressure apparatus at 4 GPa and 900 °C for 30 min. The temperature was slowly cooled before gradually releasing the pressure. This procedure resulted in around 50 mg of sample powder.

**X-ray diffraction**

The phase purity of samples was investigated by x-ray diffraction. The diffraction data were collected on a Panalytical X'pert Pro MPD diffractometer using Cu $K_{\alpha 1}$ radiation. The diffraction patterns were refined with the FULLPROF[6] software suite. The quality of the data collected on the small $Ba_2CuTeO_6$ sample was not deemed sufficient for Rietveld analysis, and therefore Le Bail full profile fitting was performed instead. Rietveld refinement was used for $Ba_2CuWO_6$. The crystal structures were visualized with VESTA.[7]

The measured x-ray diffraction patterns for $Ba_2CuTeO_6$ and $Ba_2CuWO_6$ are shown in Fig. 2. No impurity peaks are observed in $Ba_2CuTeO_6$ indicating that the material is phase pure. In the $Ba_2CuWO_6$ sample a minor (< 1%) $BaWO_4$ impurity is observed in addition to the main phase. The lattice parameters are in good agreement with literature.[8]

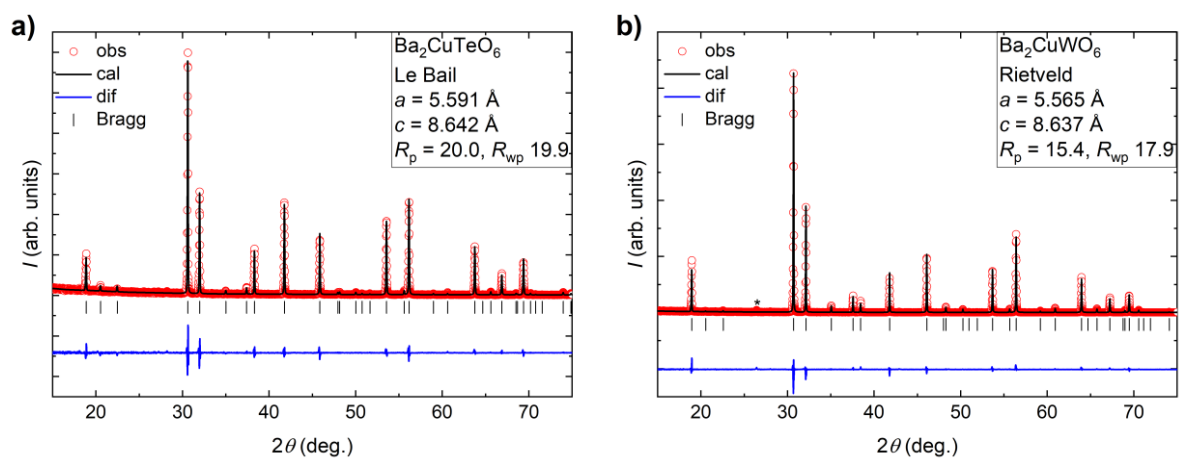

Fig. 2. X-ray diffraction patterns of (a) $Ba_2CuTeO_6$ and (b) $Ba_2CuWO_6$. The minor $BaWO_4$ impurity in $Ba_2CuWO_6$ is marked with an asterisk. Bragg positions for the space group $I4/m$ are shown.

**Magnetic measurements**

Magnetic properties were measured with a Quantum Design MPMS3 SQUID magnetometer. 120 mg of $Ba_2CuWO_6$ and 25 mg of $Ba_2CuTeO_6$ were enclosed in gelatin capsules and placed in plastic straws for measurements. DC magnetic susceptibility was measured in the temperature range 2-400 K under an applied field of 1 T in zero-field cool (ZFC) and field cool (FC) modes.